\documentclass[preprint,showpacs,preprintnumbers,amsmath,amssymb]{revtex4}

\usepackage{graphicx}
\usepackage{dcolumn}
\usepackage{bm}

\begin{document}

\title{Fiske Steps and Abrikosov Vortices in Josephson Tunnel Junctions}

\author{M. P. Lisitskiy}

\email{m.lisitskiy@cib.na.cnr.it}

\affiliation{Istituto di Cibernetica "E.Caianiello" del C. N. R.,\\
Via Campi Flegrei 34, 80078 Pozzuoli, Naples, Italy.}
\author{M. V. Fistul}
\affiliation{Theoretische Physik III, Ruhr-Universit\"at Bochum,
D-44801 Bochum, Germany}

\date{\today}

\begin{abstract}

We present a theoretical and experimental study of the Fiske
resonances in the current-voltage characteristics  of "small"
Josephson junctions with randomly distributed misaligned Abrikosov
vortices. We obtained that in the presence of Abrikosov vortices the
resonant interaction of electromagnetic waves, excited inside a
junction, with the ac Josephson current manifests itself by Fiske
steps in a current-voltage characteristics even in the absence of
external magnetic field. We found that the voltage positions of
the Fiske steps are determined by a junction size , but the Fiske
step magnitudes depend both on the density  of trapped Abrikosov
vortices and on their misalignment parameter. We measured the
magnetic field dependence of both the amplitude of the first Fiske
step and the Josephson critical current of low-dissipative small
$Nb$ based Josephson tunnel junctions with artificially introduced
Abrikosov vortices. A strong decay of the Josephson critical
current and a weak non-monotonic decrease of the first Fiske step
amplitude on the Abrikosov vortex density were observed. The
experimentally observed dependencies are well described by the
developed theory.
\end{abstract}


\maketitle

\section{\label{sec:level1}Introduction}

It is a well known that the current-voltage ($I-V$) characteristics
 of Josephson tunnel junctions can display sharp
current resonances. These interesting features, called Fiske
steps, appear as a result of resonant interaction  between ac
Josephson current and standing electromagnetic waves  that can be
excited in the junction \cite{ref1}. A long time ago the Fiske
steps have been experimentally observed in planar Josephson tunnel
junctions \cite{Fiske_Obs}. In this time the complete theoretical
description of the Fiske steps in the case of a junction with low
quality factor has been done by Kulik in \cite{Kulik1}. After that
the Fiske steps have been observed in numerous Josephson coupled
systems, e.g. Josephson junction arrays, \cite{Caputo}, stacked
artificial Josephson junctions \cite{Ustinov_stacked}, intrinsic
high-$T_c$ $Bi_2Sr_2CaCu_2O_{8+x}$ stacked superconducting tunnel
junctions \cite{Krasnov}, high-$T_c$ $YBa_2Cu_3O_{7-\delta }$
bicrystal Josephson junction \cite{Winkler}. Recently, the Fiske
steps have been investigated theoretically in $0-\pi $ Josephson
junctions \cite{Nappi} and they have been experimentally observed
in $0-\pi $ Josephson tunnel junctions with ferromagnetic barrier
\cite{Pfeiffer}.

In "small" Josephson junctions in which a  junction size is less
than the Josephson penetration depth $\lambda _j$, the spectrum of
the electromagnetic waves has a form: $\omega(k)=c_0k$, where
$c_0$ is the Swihart velocity, and the wave vectors $k_n$ are
determined by a junction´s size $W$ as $k_n=\pi n/W$, and $n=1,
2....$. The voltage positions  of $n$-th Fiske steps $V_n$ are
determined by the spectrum $\omega(k_n)$ of the electromagnetic
waves as \cite{ref1}

\begin{equation}
V_n~=~\frac{\hbar \omega(k_n)}{2e}. \label{VoltagePosition}
\end{equation}

The magnitude of Fiske steps depends strongly on the spatial
dependence of the Josephson phase difference between the
electrodes $\varphi (\vec{\rho})$, where $\vec{\rho}=\{x,y\}$ are
the coordinates in the junction plane.

In the absence of inhomogeneities a simplest way to create a
coordinate dependence of the Josephson phase difference $\varphi$
is to apply an external magnetic field parallel to the junction
plane. In this case the Josephson critical current is strongly
diminished but the magnitude of the Fiske step increases
substantially. The magnetic field dependence of Fiske steps has
been investigated in Josephson tunnel junctions with different
shapes, namely, square shaped junction \cite{Paterno}, circular
shaped junction  \cite{Nerenberg}, quatric shaped junction
\cite{quatric}. In \cite{Perezdelara} the Fiske steps have been
experimentally and theoretically investigated in the case of the
annular shaped Josephson junction both in an external magnetic
field and in a magnetic field generated by the injection current
passed along one of the annular junction electrode. However, even
in the absence of externally applied magnetic field the Fiske
resonances can appear in \emph{nonuniform Josephson junctions}.
Indeed, as was observed in Ref. \cite{Matisso},  the randomly
distributed barrier inhomogeneities (the structural fluctuations)
provide necessary conditions for generation of Fiske steps without
external magnetic field.

From the experimental point of view the nonuniform Josephson
junctions can be formed  by including artificial inhomogeneities
in the tunnel barrier \cite{ref2} or by introducing  Abrikosov
vortices \cite{ref3,ref4}. Inhomogeneities of the first type
suppress locally the Josephson tunneling, and, therefore, the
spatial variation of Josephson critical current density occurs.
These inhomogeneities may be included in the junction barrier
during the fabrication process. So a variation of some property of
the inhomogeneities (for example, their density) implies necessity
to fabricate several  Josephson junctions in which this property
varies. Obviously, the practical realization of this kind of
experiment is a rather complicated task.

On the other hand   the Abrikosov vortices
 can be easily trapped in  the junction by applying an external magnetic field
\emph{perpendicular} to the junction plane. Moreover, if pinning
centers are randomly distributed in superconducting electrodes,
the Abrikosov vortices can be trapped in the particular form of a
misaligned vortex such that the magnetic field of  the  vortex
enters and leaves the superconducting electrodes of a junction in
different points (see Fig. 1). We notice here that an extreme case
of misaligned vortices, so called "pancakes" vortices,  naturally
appears in high-$T_c$ superconductors \cite{Clem}. The misaligned
Abrikosov vortices can be considered as local magnetic
inhomogeneities leading to an additional spatial variation of the
Josephson phase difference $\varphi(\vec{\rho})$, and, hence, the
variation of  the Josephson critical current density  over a
junction area \cite{ref3,ref4,GK,Fist1}. By means of field cooling
process one can  vary the density of Abrikosov vortices in the
same junction. The possibility to trap Abrikosov vortices in a
relatively easy way offers many opportunities for experimental
studies of Josephson junctions with inhomogeneities. Thus, the
dependence of Josephson critical current on the Abrikosov vortices
density and an externally applied magnetic field have been
intensively studied \cite{ref3,ref4,GK,Fist1,JETP93,Miller} many
years ago. Recently, in Refs. \cite{Fin1,Fin2} a Josephson
junction has been used as a tool for monitoring of the position of
a single Abrikosov vortex in thermal depinning and  Lorentz force
depinning experiments. The recent experiments on a spontaneous
fluxon formation in annular Josephson junctions demonstrate the
necessity to take into account the Abrikosov vortices trapping
during the thermal quench as a competitive effect \cite{Monaco}.
In addition, a new detection principle based on the interaction of
a single gamma-photon with trapped Abrikosov vortex is proposed
for the development of a gamma -ray solid state detector with high
intrinsic detection efficiency in the energy range up to $100 ~
keV$ \cite{Lis}.

In this paper we present a theoretical study of the Fiske steps
for a "small" Josephson junctions in the presence of randomly
distributed Abrikosov vortices. Our approach is based on the
extension of a well known Kulik analysis  for low-dissipative
uniform Josephson junctions \cite{ref5} to the Josephson junctions
with inhomogeneities.  We obtain a peculiar regime where the Fiske
steps amplitude shows a weak non-monotonic decrease with the
vortex density $n_A$. We experimentally trapped different number
of Abrikosov vortices and for each vortex density $n_A$ we
measured the dependencies of the critical current $I_c$ and the
Fiske steps amplitude $I_F$ on $n_A$ and on externally applied
parallel to the junction plane magnetic field $B_{\parallel}$. A
good agreement between our theory and experimental results was
found.

The paper is organized as follows.  In section \ref{sec:level2} the theoretical model of a
Josephson junction with randomly distributed misaligned Abrikosov vortices is
 presented. In section \ref{sec:level3} by making use of a generic approach
elaborated in the Ref. \cite{ref5}, we  calculate the Fiske
resonances in inhomogeneous Josephson junctions, and in the
section \ref{sec:level4} the dependence of the Fiske steps
amplitude $I_F$ on the density of Abrikosov vortices $n_A$ is
analyzed. Section \ref{sec:level5} is dedicated to
 the experimental details, namely, the experimental setup, the sample
description and the procedure to  measure the Josephson critical
current $I_c$ and the amplitude of first Fiske step $I_F$. In
section \ref{sec:level6} we present the experimental results and
the comparison with the theory. Section \ref{sec:level7} provides
conclusions.

\section{\label{sec:level2} Model of a Josephson junction with randomly distributed
misaligned Abrikosov vortices}

We consider a small, i.e. $W<\lambda_J$, Josephson junction in the
presence of randomly distributed pinned misaligned Abrikosov
vortices (here, $W$ is the size of a Josephson junction, and
$\lambda_J$ is the Josephson penetration depth).  A magnetic field of a misaligned
Abrikosov vortex enters and leaves the superconducting electrodes
of a junction in different points. The distance between these
points determines the misalignment length $\delta$ (see Fig. 1).
%
%
\begin{figure}
\includegraphics[width=8cm]{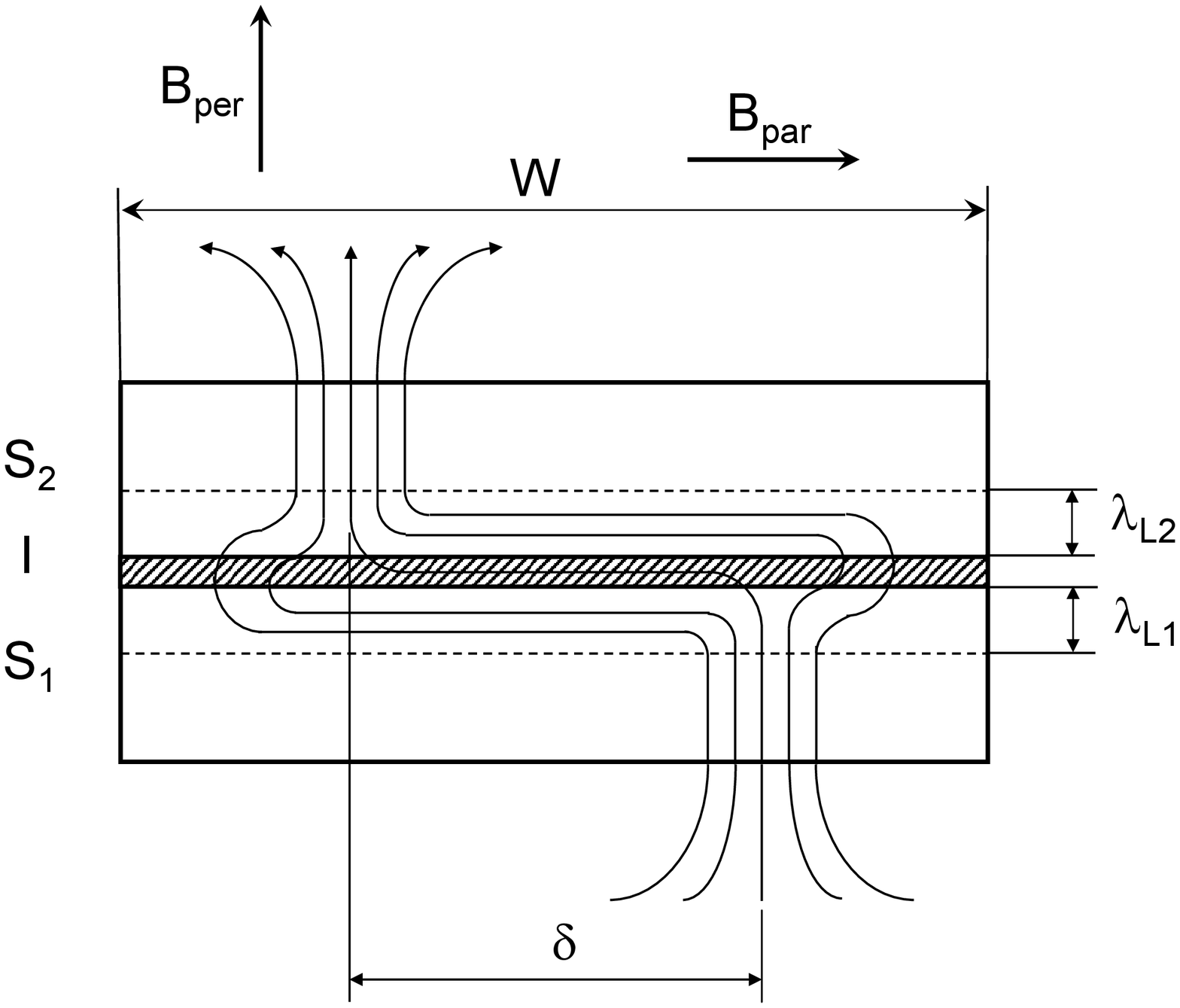}
\caption{\label{fig1} Schematic cross-section of a Josephson
tunnel junction with a misaligned Abrikosov vortex. $\delta$ is
the misalignment length. $\lambda _{L_1}$ and $\lambda _{L_2}$ are
the London penetration depths of the superconductors forming the
junction. The orientation of the two magnetic field, $B_{\perp}$
and $B_{\parallel}$, is provided. }
\end{figure}
%
%
Under these conditions the Abrikosov vortex contribution
$\varphi_V(\vec{\rho})$ to the total Josephson phase difference $\varphi
(\vec{\rho},t)$ randomly depends  on the position $\vec{\rho} $ in
the plane of the junction. The particular form of a single Abrikosov vortex
contribution has been obtained previously in Refs.
\cite{GK,Fist1}.

The dynamics of a Josephson junction with randomly distributed Abrikosov vortices
is described by an inhomogeneous
sine-Gordon equation \cite{ref1,ref6,FG}:

$$
\frac{\partial^2 \varphi_1(\vec{\rho},t)}{\partial\vec{\rho}^2} -
 \frac1{c_0^2} \frac{\partial^2\varphi_1(\vec{\rho},t)}{\partial t^2}
- \frac{\gamma}{c_0^2}
\frac{\partial\varphi_1(\vec{\rho},t)}{\partial t }
$$
\begin{equation}
{=}\lambda_J^{-2}\sin [\varphi_V( \vec{\rho} ) + k_0x+\omega_0t +
\varphi_1( \vec{\rho}, t )], \label{singordon}
\end{equation}
where
$$
k_0= \frac{2 \pi B_{\parallel}( \lambda_{L1}+ \lambda_{L2})}{\Phi_0},~
\omega_0 = \frac{2e{V}}{ \hbar}.
$$
Here $V$ is the external dc voltage, $c_0$ is the velocity of
Swihart waves, $B_{\parallel}$ is an external magnetic field applied
parallel to the $y$-axis, $\lambda _{L_1}$ and $\lambda _{L_2}$
are the London penetration depths of the superconductors forming
the junction, the parameter $\gamma$ determines the decay of the
electromagnetic waves in the junction and depends on the
quasiparticle resistance of the junction in a subgap region,
$\Phi_0$ is the flux quantum .

\section{\label{sec:level3}Fiske resonances in inhomogeneous Josephson tunnel
junctions: theory}

Our theoretical analysis  is based on the extension of the Kulik
approach developed for small low-dissipative Josephson  junctions with
$\gamma<<\omega_0$,   \cite{ref1,ref5}. Thus,
we present $\varphi_1(\vec{\rho},t)$ in the following form

\begin{equation}
 \varphi_1(\vec{\rho},t)=\sum_n\Re e \{a_n e^{-ik_nx-i\omega_0t} \},\label{Josphase}
\end{equation}%
where the wave vectors of electromagnetic waves that can be
excited in the junction, i.e.  $k_n=\pi n/W$, are determined by
the junction's geometry. We note that in the particular case when
the parameter $\gamma$ is not so small, Fiske resonances overlap
each other, and a single Eck peak is formed in the $I-V$
characteristic of a Josephson junction \cite{ref1}. Such feature
has been theoretically analyzed also for Josephson junctions with
randomly distributed inhomogeneities \cite{FG}.

As the parameter $\gamma<<\omega_0$ the Fiske resonances are well
separated, and for the analysis of $n$-th Fiske step we can choose
a single wave vector $k_n$ in the Eq. (\ref{Josphase}).
Substituting Eq.(\ref{Josphase}) in Eq.(\ref{singordon}) and using
the condition $\hbar \omega_0=c_0k_n$, i.e. (\ref{VoltagePosition}),
we obtain the nonlinear equation for the amplitude of Swihart
electromagnetic waves as

\begin{equation}
a_n=\frac{c_0^2}{\lambda_J^2 \gamma \omega_0} [J_0^2
(\frac{a_n}{2}) + J_1^2( \frac{a_n}{2}) ] Re [\int \frac{d^2
\vec{\rho}}{S} ~e^{ i \varphi_V (\vec{\rho})+ik_0x-ik_nx}],
\label{nonlinampl}
\end{equation}

where  $J_0$ and $J_1$ are the Bessel functions, $S$ is the junction area.

Similarly we obtain the expression for the amplitude of $n$-th
Fiske step
\begin{equation}
I^{(n)}_F ~=~ j_c J_0 (\frac{a_n}{2}) J_1 (\frac{a_n}{2}) Re [\int
d^2 \vec{\rho} ~e^{ i \varphi_V (\vec{\rho})+ik_0x-ik_nx}]
~.\label{Joscurrent-gen}
\end{equation}
where $j_c$ is the Josephson critical current density in the
absence of vortices.

\section{\label{sec:level4}Critical current and Fiske resonances: the dependence
on the Abrikosov vortices density}

Since the  contribution of Abrikosov vortices $\varphi_V
(\vec{\rho})$ to the Josephson phase difference is a random
function of the coordinate $\vec{\rho}$, we  obtain the averaged
quantities only. First we define the correlation area for
Abrikosov vortices as
\begin{equation}
\sigma(k)=<\int d^2\vec{\rho}
e^{i[\varphi_V(\vec{\rho})-\varphi_V(0)]+ikx}>, \label{Corrarea}
\end{equation}
where the sign $<....>$ means the averaged value over the random
positions of Abrikosov vortices. The correlation area $\sigma(k)$
depends on the density  of misaligned Abrikosov vortices $n_A$,
the misalignment length $\delta$, and the wave vector $k$ that, in
turn, is determined by an externally applied magnetic field
parallel to the junction plane $B_{\parallel}$. For the particular
case of a circular junction the $\sigma(k)$ has been found in
\cite{ref3,ref4,Fist1}, and in the absence of $B_{\parallel}$

\begin{equation}
\sigma  ~\simeq~ S \exp (-\pi n_A {\delta^2} \ln (2L/ \delta)) ~,
\label{correlationarea}
\end{equation}
where $L$ is the radius of the circular junction. It has been
obtained in \cite{ref3,ref4,Fist1} that the critical current of
the Josephson junction with randomly distributed misaligned
Abrikosov vortices is determined by the correlation area
$\sigma(k)$ as

\begin{equation}
I_c ~=~ j_c \sqrt{ \sigma(k)  S } .\label{Criticalcurrent-1}
\end{equation}
Thus, in the absence of $B_{\parallel}$ as the critical current
reaches the maximum, we obtain a strong decrease of the critical
current $I_c$  with the Abrikosov vortices density $n_A$:

\begin{equation}
ln{\frac{I_c^{max}}{I_{c0}}}=\frac{n_A
\pi{\delta}^2}{2}ln{\frac{\delta}{2L}}, \label{Criticalcurrent-2}
\end{equation}
where $I_{c0}$ is the critical current in the absence of Abrikosov vortices.

Next, we turn to the analysis of  Eqs. (\ref{nonlinampl}) and
(\ref{Joscurrent-gen}) determining the amplitudes of Fiske
resonances. First, we notice that even in the absence of $k_0$ (or
$B_{\parallel}$) the amplitudes of Fiske resonances $I^{(n)}_F$ are not
zero. However, the magnitudes of Fiske steps can be increased by
tuning of $B_{\parallel}$. The maximum values of $I^{(n)}_F$ reach as
the condition $k_n=k_0$ is satisfied. In the following we will
assume that the correlation area $\sigma$ is small, i.e. $\sigma
\ll S$. Using this assumption we express the maximum amplitudes of
Fiske resonances $I^{(n)}_F$ through a single parameter
$\sigma(0)=\sigma$. Indeed, in the limit as $\sigma $ is small or
more precisely $\sigma  ~\ll~ S (\frac{ \lambda_J^2 \gamma
\omega_0 } { c_0^2} )^2~=~\sigma^*$, the amplitudes $a_n$ are
small, and expanding the Bessel function over a small argument
$a_n$ we obtain
\begin{equation}
I^{(1)}_{F} ~=\frac{j_c \sigma}{4}\sqrt{\frac{S}{\sigma^*}}~=~j_c
\sigma c_0^2 (4 \lambda_j^2 \gamma \omega_0)^{-1} ,~~~ \sigma \ll
\sigma^* \label{Fiske-small}
\end{equation}
In this regime  $I^{(1)}_{F}$ is proportional to $\sigma$ and
displays a strong exponential decrease with $n_A$.

In the opposite limit $\sigma  ~\geq~ \sigma^*$, the amplitude of
electromagnetic waves becomes large but the oscillations of Bessel
functions are strongly suppressed. Therefore, the amplitude of
$I^{(1)}_{F}$ is still small. The averaged value of $I^{(1)}_{F}$
can be also expressed through a single parameter $\sigma$ as
\begin{equation}
I^{(1)}_{F} ~=(\frac{32}{\pi})^{1/3} j_c S (\frac{\sigma^2
\sigma^*}{S^3})^{1/6})e^{-\frac{3}{8}[\frac{4\sigma}{\pi\sigma^*}]^{1/3}},~~\sigma
~\gg~ \sigma^*~\label{IF-large}
\end{equation}

In this limit, the amplitude of Fiske resonance weakly depends on
the density of Abrikosov vortices, displaying a small maximum on
$\sigma~\simeq~\sigma^*$. Notice here, that this regime can be
easily realized for low-dissipative junctions as
$\gamma<<\omega_0$.

\section{\label{sec:level5}Josephson junctions with misaligned Abrikosov vortices:
experimental setup and measurements}

The experiments were carried out on the $Nb-AlO_x-Nb$ Josephson
tunnel junction of square geometry and dimensions of
$50~\times~50~{\mu m}^2$. Details of the fabrication process are
reported in \cite{ref6,ref7}.

The junction was mounted on the sample holder inside a closed
copper box in order to prevent the influence of the
electromagnetic noise. The box was immersed in a liquid helium
bath, so all measurements were done at a stable temperature $T=4.2
K$. In order to heat the junction above the superconducting
transition temperature $T_c$, a SMD 100 Ohm resistor was used as a
heater mounted in contact with the sample holder. A special coil
system containing two Helmholtz copper coils was used in order to
generate both the magnetic fields perpendicular to the junction
plane $B_{\perp}$ and the magnetic field  parallel  to the
junction plane  $B_{\parallel}$. The shielding of the sample from
the Earth magnetic field and from the electromagnetic noise was
reached by using a three level $\mu$-metal shielding system and a
one level $Al$-shield.


\begin{figure}
\includegraphics[width=8cm]{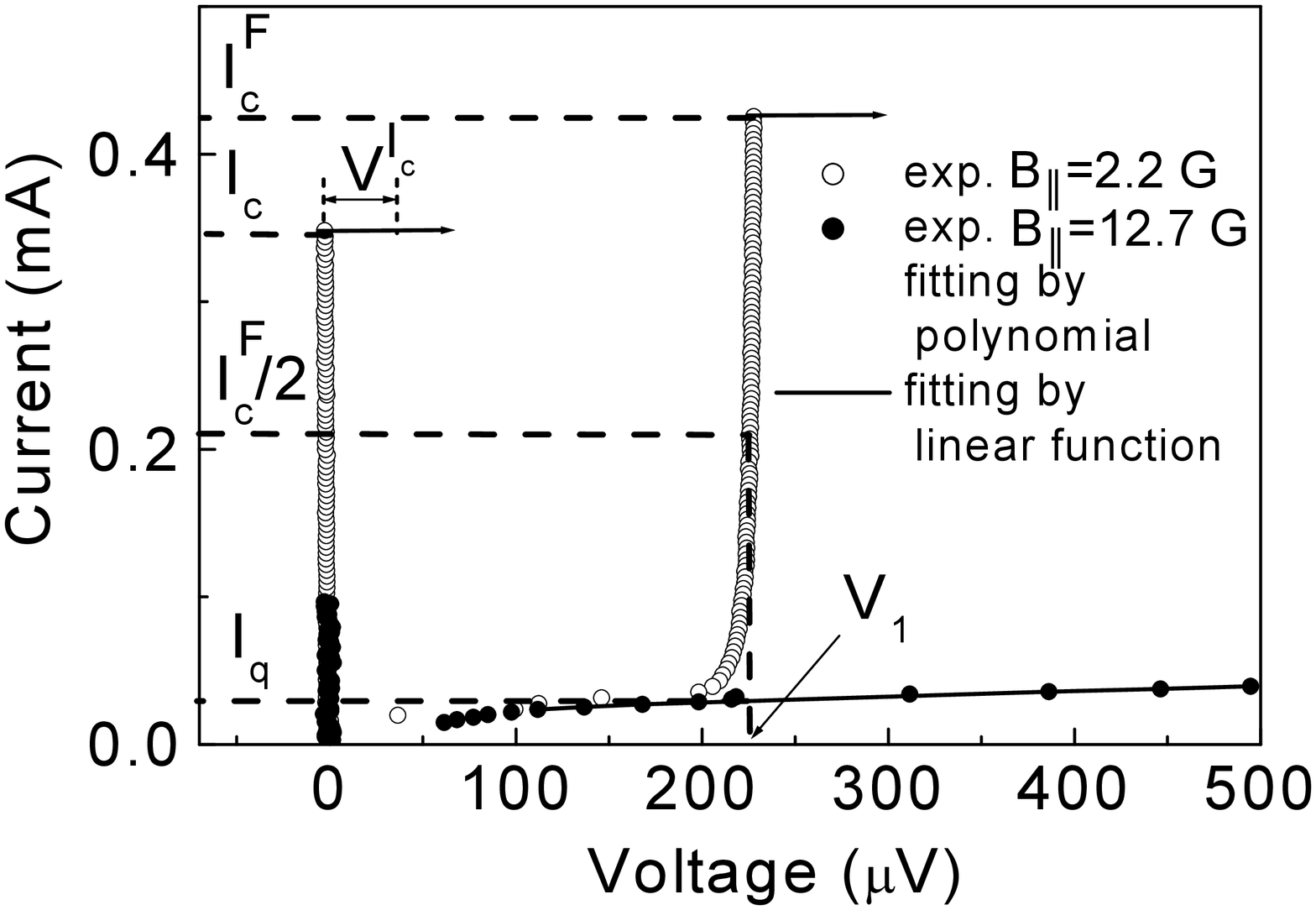}
\caption{\label{fig2}$I-V$ curves recorded at $B_{\parallel}=2.2 G$
when the first Fiske step is maximized (open circles) and at
$B_{\parallel}=12.7 G$ when the Fiske step is suppressed (closed circles,
). The method allowing to extract both the amplitude of the first
Fiske step, $I^{(1)}_F$, and its voltage position, $V_1$ is shown.
The voltage position is obtained by the dashed line which is
parallel to the current axis and intersects the Fiske step branch
at $I=I_c^{F}/2$. The amplitude of the first Fiske step is
$I^{(1)}_F=I_c^{F}-I_q$ where $I_q$ is the quasiparticle
current at voltage $V_1$. $V^{I_c}$ is the threshold voltage to
define the Josephson critical current.
  }
\end{figure}

The accurate measurements both the Josephson critical current
$I_c$, and the amplitude of the first Fiske step $I_F^{(1)}$ were
achieved by a careful shielding and by inserting cold low pass
filter, mounted just near the sample. First, we measured the $I-V$ curves of
the junction without
trapped Abrikosov vortices. We  tuned the amplitudes of Fiske resonances
by an externally applied magnetic field $B_{\parallel}$. Fig.2 shows the
typical $I-V$ curves where  the amplitude of the first Fiske
step was maximized (open circles) and suppressed (closed circles). The first $I-V$
curve was used in order to obtain
the voltage position of the Fiske step $V_1$,, while the second
one was utilized for experimental estimation of the quasiparticle
current in the sub-gap region.

The method to determine the  voltage
position $V_1$ of the first Fiske step  is shown in Fig.2. We note that the voltage
position can also be  estimated as the voltage at current
$I_c^{F}$ (see Fig.2) but the difference between this method
and the one represented in Fig.2 is negligible.

The experimental  technique to measure the Josephson critical current
$I_c$ was as follows. The current $I$ was increased linearly in
time from zero value, and the junction voltage $V$ was
simultaneously recorded. When the current reached the Josephson
critical current value, a non zero junction voltage appeared.
Under a condition that the junction voltage equals to the
threshold voltage $V^{I_c}$  (criterion of the Josephson critical
current, see Fig.2), the current value was recorded and the
current returned back to zero and then the next measuring cycle
started again. The current sweep frequency was 10~Hz and the value
of $V^{I_c}$ was of $10~\mu V$.

A  cycle to measure the amplitude of the first Fiske step was
similar to the Josephson critical current measurements but the
starting value of the current was chosen to satisfy the condition
that the junction voltage was greater then zero and less then
voltage position of the first Fiske step. From this starting point
the current was increased, the first Fiske step branch was
recorded and the maximum current at the first Fiske step branch
$I_c^{F}$
 was defined as the
current at which the junction voltage became greater then
$V_1+V^{I_c}$ (see Fig.2). The amplitude of the first Fiske step
was determined by substracting the quasiparticle current at Fiske
step voltage $I_q$ from the switching current at Fiske branch
$I_c^{F}$. Both Josephson critical current and amplitude of the
first Fiske step for a given value of parallel magnetic field
$B_{\parallel}$ were obtained. The $I_c(B_{\parallel})$ and $I^{(1)}_F(B_{\parallel})$ dependencies
were measured for  both polarities of
the bias current.

The experimental cycle for the evaluation of the influence of
Abrikosov vortices
on the Fiske step amplitude consisted of the following steps:\\
1) The sample was heated to $T>T_c$;\\
2) A magnetic field $B_{\perp}$ of fixed value was applied;\\
3) The junction was cooled in the field $B_{\perp}$ up to $T=4.2~K$, i.e. immersed in the liquid helium
at normal pressure;\\
4) The magnetic field $B_{\perp}$ was turned off;\\
5) The amplitude of the first Fiske step was maximized by the parallel magnetic field $B_{\parallel}$ and the $I- V$ curve with the
first Fiske step branch was
measured in order to obtain the voltage position of the first Fiske step;\\
6) The $I_c(B_{\parallel})$ curve was measured;\\
7) The $I_c^{F_1}(B_{\parallel})$ curve was measured;\\
8) Both the Josephson
critical current and the amplitude of the first Fiske step, were suppressed by a
suitable parallel magnetic field $B_{\parallel}$ and the $I -V$ curve was recorded in order to measure
the quasiparticle current at $V=V_{1}$.

After the field cooling process the Abrikosov vortex density $n_A$
can be estimated as

\begin{equation}
n_A\simeq \frac{B_{\perp}}{\Phi_0}. \label{Vortexdensity}
\end{equation}

\section{\label{sec:level6}Experimental Results and Discussion}

\subsection{\label{sec:level6a}Junction Parameters}

The measurements  of the $I_c(B_{\parallel})$ dependence without
trapped Abrikosov vortices were performed at $T=4.2~K$ (Fig.3, open circles). The
Josephson penetration depth $\lambda_j$ was estimated to be
about 54~$\mu m$. The experimental $I_c(B_{\parallel})$ curve is in
a good agreement with the theoretical one which was calculated for
a "small" square junction by the Fraunhofer formula \cite{ref1}

\begin{equation}
I_c(B_{\parallel})=j_c{W^2}\left|\frac{sin(\frac{\pi{B_{\parallel}}Wd}{\Phi_0})}
{\frac{\pi{B_{\parallel}}Wd}{\Phi_0}}\right|~, \label{Fraunhofer}
\end{equation}

where $j_c$ is the Josephson critical current density, $W$ is the
dimension of the junction, $d=\lambda_{L_1}+\lambda_{L_2}+t$ and
$\lambda_{L_1}$ and $\lambda_{L_2}$ are the effective London
penetration depths of the superconductors forming the junction,
and $t$ is the  barrier thickness. The theoretical curve calculated by
Eq.(\ref{Fraunhofer}) is reported in Fig.3 by solid line.
This good agreement between experimental and theoretical results
confirms the uniform distribution of the Josephson current density
over the junction area.
%
%
\begin{figure}
\includegraphics[width=8cm]{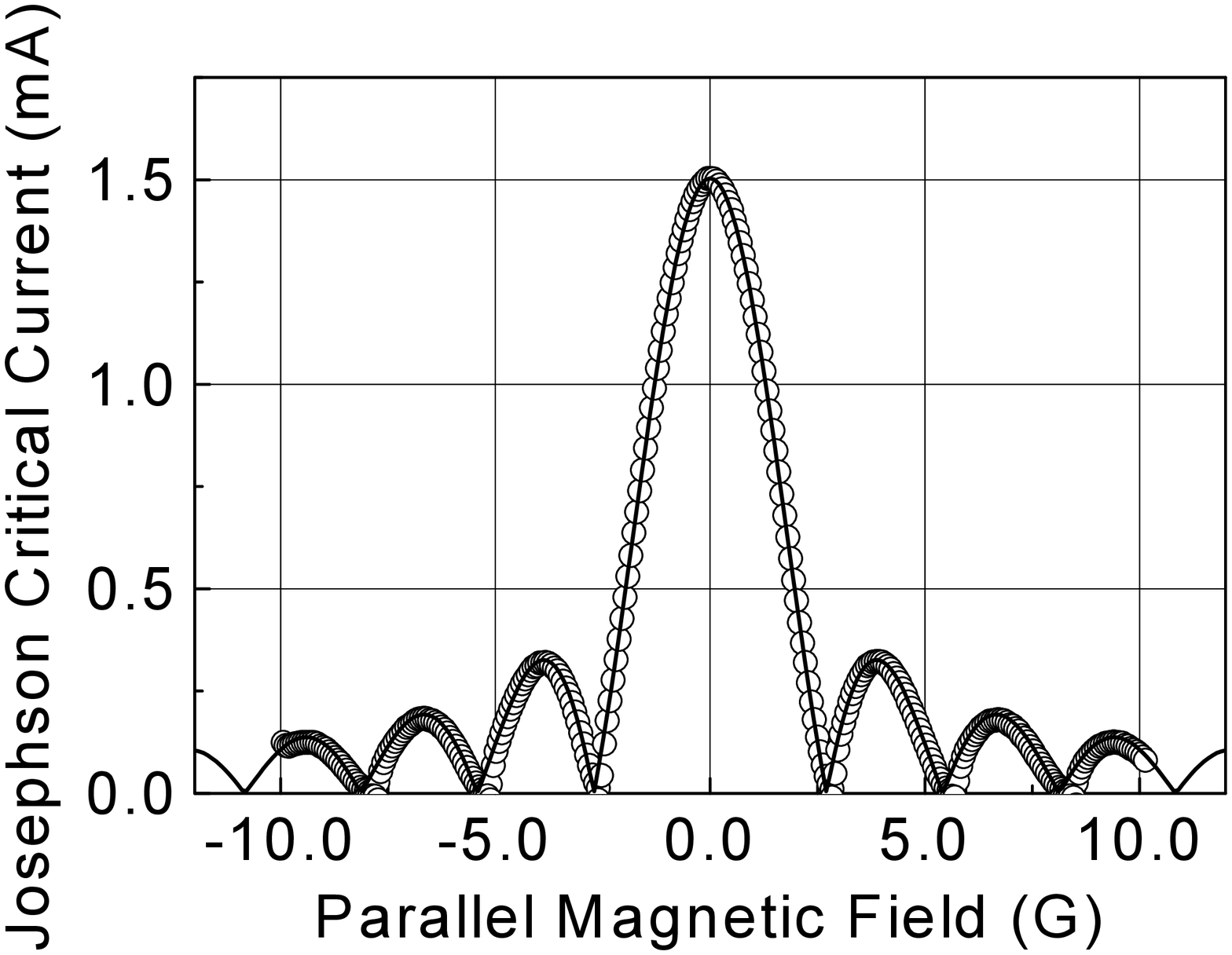}
\caption{\label{fig3} The experimental $I_c(B_{\parallel})$ curve
(open circles) and theoretical fit by Eq.(10) (solid line). The
trapped Abrikosov vortices are absent.
  }
\end{figure}

\subsection{\label{sec:level6b}Magnetic field dependence of the first
Fiske Step without Abrikosov vortices}

In the absence of trapped Abrikosov vortices the value of voltage position of
the first Fiske step $V_1$ was
determined from the $I-V$ curve at $T=4.2~K$ when the Fiske
step amplitude was maximized by magnetic field $B_{\parallel}$ (Fig.2, open
circles). Using this value of $V_1$ and Eq.(1), we calculated the
resonant frequency $\omega_0$ and Swihart velocity $c_0$. The
values of $V_1$, $\omega_0$ and $c_0$ are
given in Table 1.

\begin{table}
\caption{\label{tab:table1} Experimental values of
the voltage position of the first Fiske step $V_1$, the Swihart
velocity, $c_0$, and the resonant frequency, $\omega_0$, obtained from Eq.(1).}
\begin{ruledtabular}
\begin{tabular}{ccccc}
$V_1$ ($\mu V$) & $c_0$ ($m/s$) & $\omega_0$ ($s^{-1}$) \\
\hline
226.3 & $1.1\times10^7$ & $6.9\times10^{11}$  \\
\end{tabular}
\end{ruledtabular}
\end{table}

The $I^{(1)}_F(B_{\parallel})$ curve  measured after cooling
the junction in a zero magnetic field $B_{\perp}$ (open circles), is
shown in Fig. 4. The theoretical dependence for the first
Fiske step amplitude on field $B_{\parallel}$ was numerically computed
from Eq. (\ref{nonlinampl}) and Eq. (\ref{Joscurrent-gen}) in the
 absence of Abrikosov vortices. The value of  the parameter
 $\gamma=1.1~\times~10^{10} ~s^{-1}$ provides a best fitting of
 experimental data. Similar value of $\gamma$ can be extracted
 from the $I$-$V$ curves where the Fiske resonances were suppressed.
The agreement with the theoretical analysis is excellent even for
the second and third lobes. In the Ref. \cite{ref5}  a similar theoretical analysis of the
Fiske resonances of a junction without Abrikosov vortices has been carried out. However, in the
equation determining the amplitude $a$ of electromagnetic waves the term of $J_1$ was
omitted. Taking into account this term one can get a better agreement with experimental data.

\begin{figure}
\includegraphics[width=8cm]{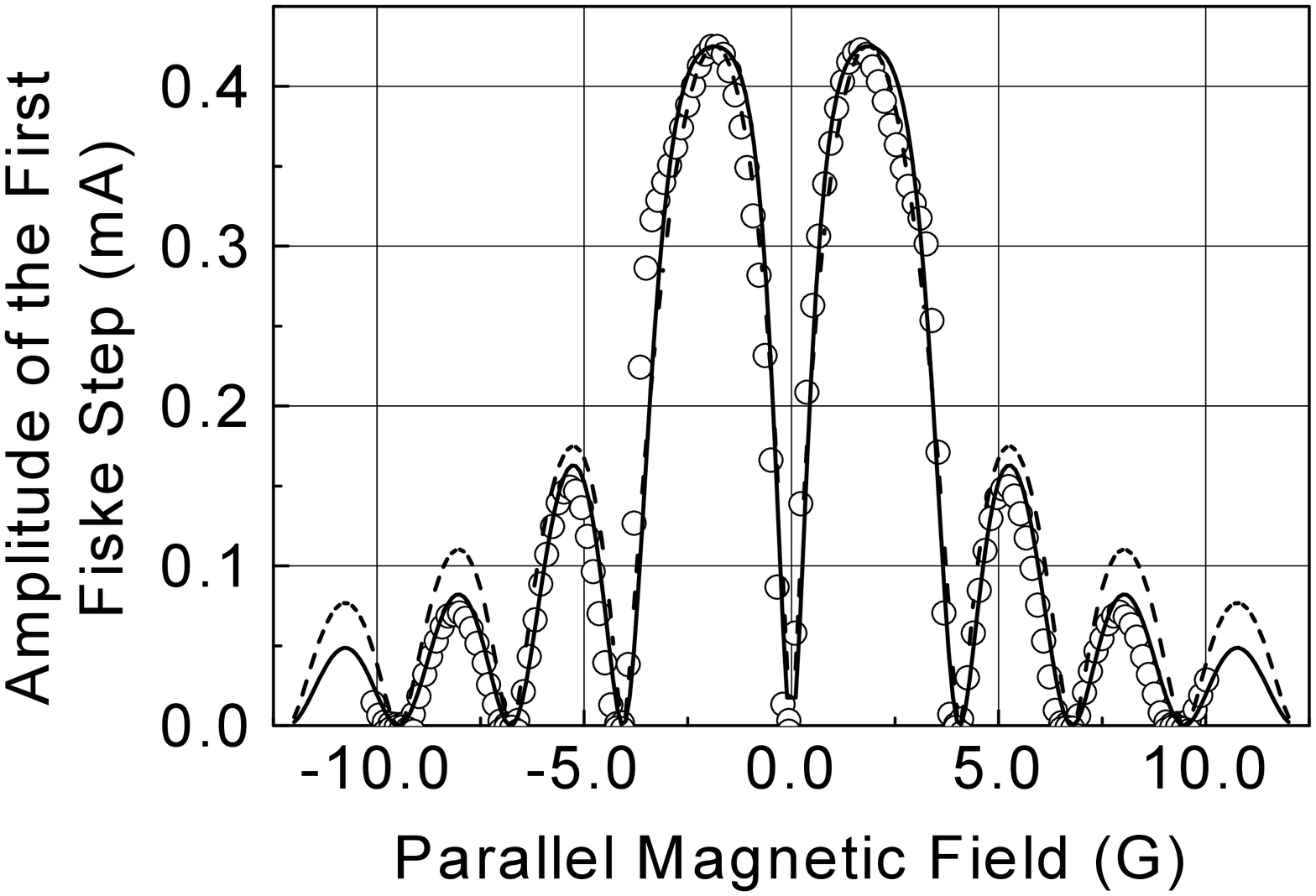}
\caption{\label{fig4} Experimental $I^{(1)}_F(B_{\parallel})$
curve for the junction without trapped Abrikosov (open circles).
Solid line represents the theoretical dependence calculated by
Eq.(\ref{nonlinampl}) and Eq.(\ref{Joscurrent-gen}) for
$\gamma=1.1\times10^{10} s^{-1}$. Dashed line is the result of
calculation without $J_1^2(\frac{a}{2})$ term at
$\gamma=4.3\times10^{9}~s^{-1}$ according to the original
theoretical analysis of \cite{ref5}.}
\end{figure}

\subsection{\label{sec:level6c}Josephson critical current of a junction with trapped
Abrikosov vortices}

\begin{figure*}
\includegraphics[width=12cm]{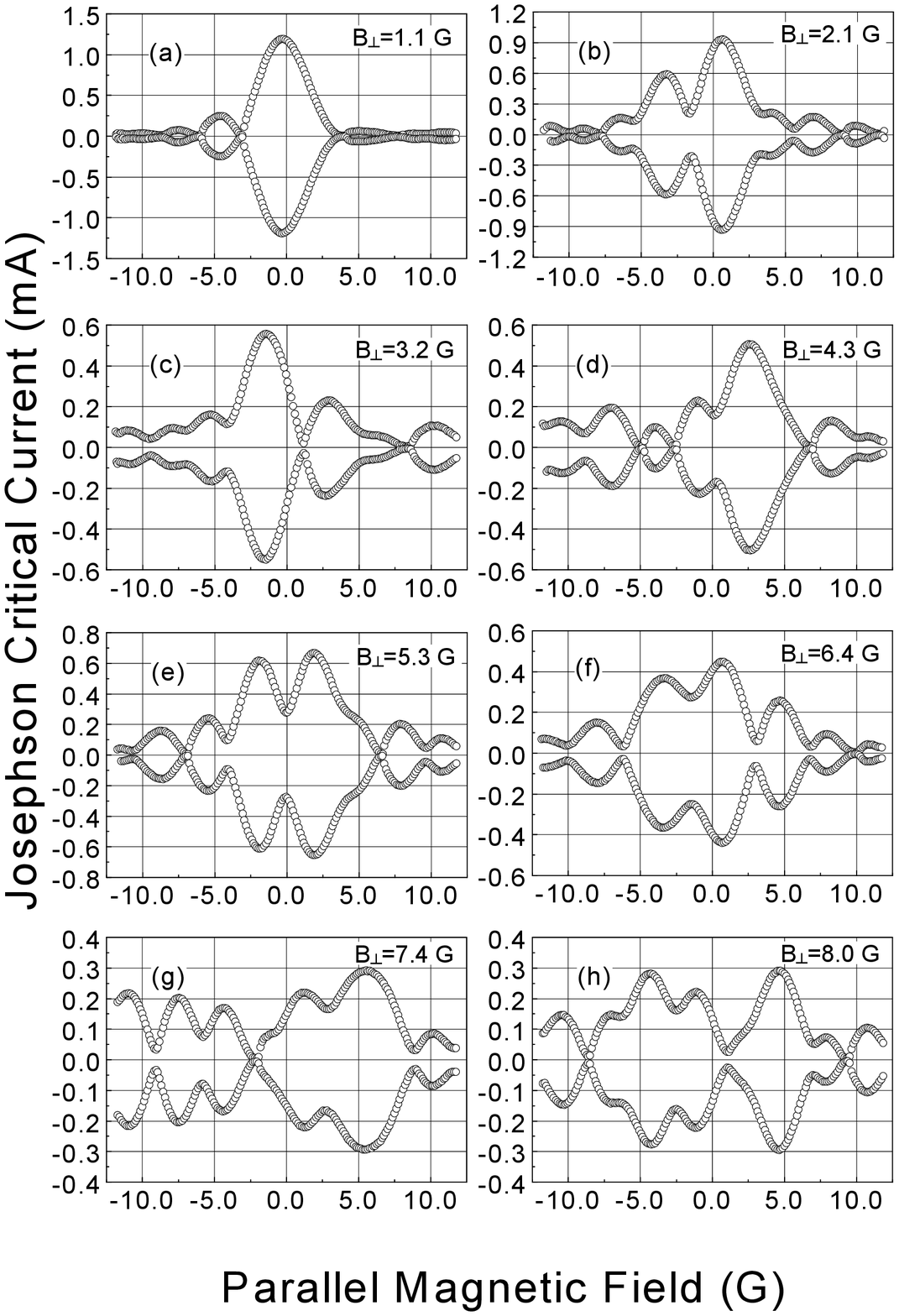}
\caption{\label{fig5} $I_c(B_{\parallel})$ curves measured after cooling of the junction  in
perpendicular magnetic field $B_{\perp}$ of various values.
  }
\end{figure*}
%
\begin{figure*}
\includegraphics[width=12cm]{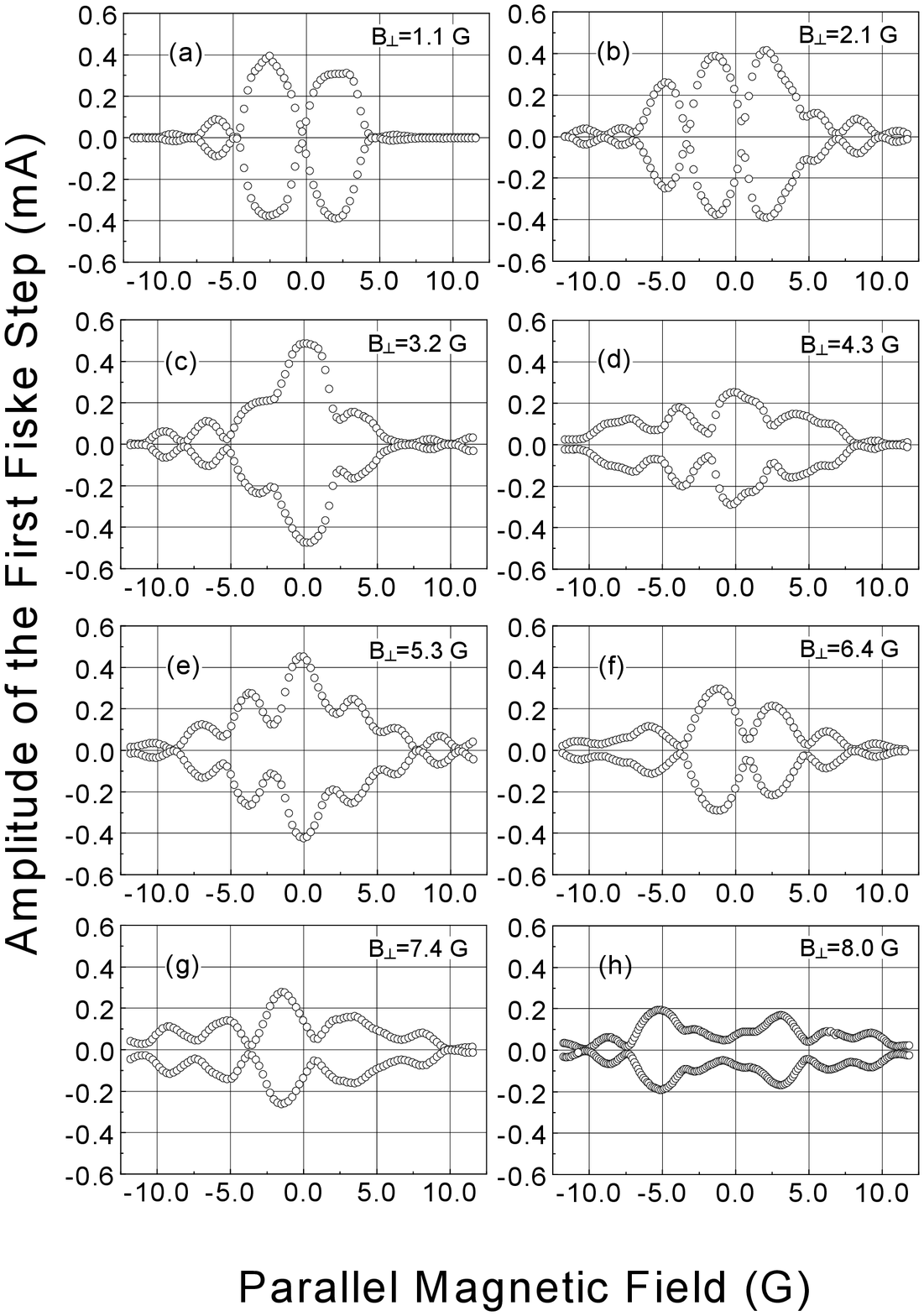}
\caption{\label{fig6} $I^{(1)}_F(B_{\parallel})$ curves measured
after the junction was cooled in perpendicular magnetic field $H_
{per}$ of various values. }
\end{figure*}
Fig.5 shows $I_c(B_{\parallel})$ curves measured after cooling
of the sample  in different  perpendicular magnetic fields $B_{\perp}$.  We
obtained from each $I_c(B_{\parallel})$ curve the maximum values of $I_c$ for both polarities,
$I_c^{+max}$ and $I_c^{-max}$,
and then we selected the maximum value between  $I_c^{+max}$ and
$I_c^{-max}$ denoting it as $I_c^{max}$.
Similar to \cite{ref3,ref4}, we found the deviation of the
$I_c(B_{\parallel})$ curve from a "vortex-free" Fraunhofer-like
curve (Fig.3, open circles) and we observed a strong suppression
of $I_c^{max}$. For each  value of $B_{\perp}$ the vortex density
$n_A$ was calculated by using Eq.(\ref{Vortexdensity}). The
dependence of the maximum Josephson critical current $I_c^{max}$
on the density of trapped Abrikosov vortices $n_A$ is shown in
Fig. 7 by open squares. It is well known that during the field
cooling process, misaligned Abrikosov vortices are randomly
trapped over the junction area (see Fig.1).
Eq.(\ref{Criticalcurrent-2}) describes the dependence of
$I_c^{max}$ both on the vortex density $n_A$ and on the
misalignment length $\delta$ in case of a circular junction.
Taking $L$ to be the radius of a circular junction with an area
equal $50~\times~50~{\mu m}^2$, we used
Eq.(\ref{Criticalcurrent-2}) and Eq.(\ref{Vortexdensity}) in order
to extract the value of $\delta$ for each value of vortex density.
The average value of misalignment parameter over the all vortex
densities is $\overline{\delta}=0.8~\pm~0.1 ~\mu m$.

\subsection{\label{sec:level6d}First Fiske step in the presence of  trapped
Abrikosov vortices}

We observed that the presence of trapped Abrikosov vortices leads
to a great deviation of the $I^{(1)}_F(B_{\parallel})$ curves from
the "vortex- free" one shown in Fig. 4 by open circles. These
curves are shown in Fig. 6. The presence of trapped vortices did
not lead to the variation of the voltage positions of  Fiske
resonance. This result indicates that the trapped Abrikosov
vortices did not change the geometrical conditions for appearance
of the first Fiske step and only influenced on the magnitude of
$I_c^{F_1}$ by additional spatial variation of the Josephson phase
difference $\varphi (\vec{\rho},t)$.

Next, we notice that a non-zero amplitude of the first Fiske step
at $B_{\parallel}=0$ appeared as Abrikosov vortices were trapped.
However, by tuning the parallel magnetic field we could
substantially increase the amplitude of the Fiske step. Using the
same procedure which was used to obtain the $I_c^{max}$ from the
$I_c(B_{\parallel})$ curves (see subsection \ref{sec:level6c} ),
we found the maximum amplitude of the first Fiske step
$I^{(1),max}_F$  for each value of $B_{\perp}$. The experimental
dependence of $I^{(1),max}_F$ on the Abrikosov vortex density
$n_A$ is shown in Fig. 7 (open circles). In contrast to the
$I_c^{max}(n_A)$ dependence, the weak nonmonotonic  decrease of
the magnitude of $I^{(1),max}_F$ with
 $n_A$ was observed. Such behavior of the $I_c^{F_1,max}(n_A)$
dependence indicates that the condition of $\sigma  ~\geq~
\sigma^*$ should be applied for our sample. Indeed, this
conclusion is confirmed by estimating the values of both $\sigma$
and $\sigma^*$ for our junction.
%
%
\begin{figure}
\includegraphics[width=8cm]{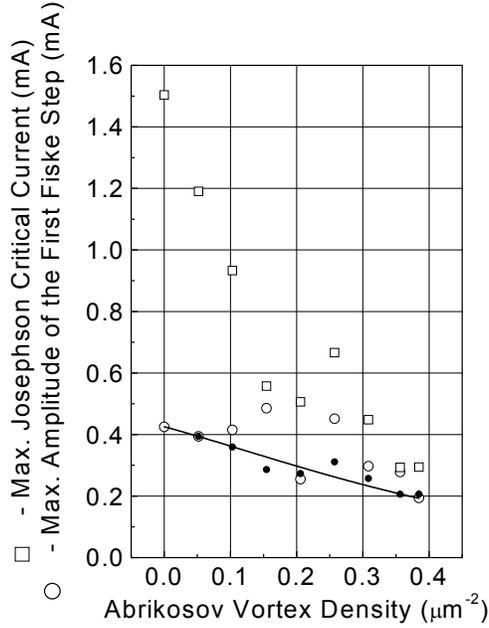}
\caption{\label{fig7} Open squares: Experimental dependence of the
maximum Josephson critical current $I_c^{max}$ on the density of
trapped Abrikosov vortices $n$; Open circles:   Experimental
dependence of the maximum amplitude of the first Fiske step
$I^{(1),max}_F$ on the density of trapped Abrikosov vortices $n$.
Solid line is the theoretical $I^{(1),max}_F(n_A)$ curve,
calculated by Eqs.(\ref{IF-large}) and (\ref{correlationarea}).
The values of $\gamma=2.2\times10^{10}~s^{-1}$ and a  mean
misalignment parameter over all vortex densities
$\overline{\delta}=0.8 \mu m$ were used; Full circles represent
the theoretical $I_c^{F_1,max}(n_A)$ dependence obtained by
Eqs.(\ref{IF-large}) and (\ref{Criticalcurrent-1}).}
\end{figure}

In order to estimate the critical value of $\sigma^*$, we have to
obtain the dissipation parameter $\gamma$. This parameter
determines the decay of the electromagnetic waves in the junction,
and it can be obtained as
 $\gamma \simeq {(R_qC)}^{-1}$ where $C$ is the
capacitance of a junction per unit area, $R_q$ is the
quasiparticle tunneling junction resistance at voltage $V=V_1$ per
unit area \cite{ref1}. Trapped Abrikosov vortices increase the
quasiparticle current by virtue of their normal cores
\cite{GK,ref8} and, the value of $\gamma$ has to also increase.
Therefore the value of $\gamma=1.1~\times~10^{10} ~s^{-1}$, obtained for
the junction in the absence of Abrikosov vortices, was not
convenient for the case of trapped Abrikosov vortices. We used
$\gamma$ as a fitting parameter in order to reach the good
agreement between the experimental value of $I^{(1)}_F$ recorded
at $B_{\perp}=1.1~G$ and the theoretical value calculated by using
Eq. (\ref{nonlinampl}) and Eq. (\ref{Joscurrent-gen}) for the same
magnetic field. We obtain $\gamma=2.2~\times~10^{10}~s^{-1}$ from this
fitting procedure and the parameter $\sigma^*=3.2~\times~10^2
~\mu m^2$ was estimated. It is important to note that the quality
factor $Q=\omega_0/\gamma$ of the Fiske resonance is equal to 31,
such that a crucial assumption of low dissipative junctions is
valid for our junctions.

Using Eq. (\ref{correlationarea}) we calculated the correlation
area $\sigma$ for each value of density $n_A$, and we obtained
that for all values of $B_{\perp}$, the correlation area
$\sigma\ge\sigma^*$, and therefore, the Eq. (\ref{IF-large}) is
valid. The theoretical dependence of $I^{(1)}_F$ on $n_A$ is shown
in Fig. 7 by solid line.

The experimental dependence $I^{(1),max}_F(n_A)$  agrees
qualitatively with  the behavior of theoretical curve but several
experimental points disagree with our theoretical predictions. One
of possible reason of this discrepancy is the variation of the
misalignment length $\delta$ for different magnetic fields
$B_{\perp}$. In order to check this hypothesis we extracted
$\sigma$ from the experimental values of $I_c^{max}$ by using Eq.
(\ref{Criticalcurrent-1}) for different values of $n_A$.
Substituting the obtained values of $\sigma$ into
Eq.(\ref{IF-large}) we calculated $I^{(1),max}_F$ for each
experimental values of $n_A$. The result of such procedure is
shown in Fig.7 by closed circles. By making use of this procedure
the agreement between the theoretical dependence and experiment
becomes better. So we can conclude that the variation of $\delta$ strongly
influences the maximum amplitude of the first Fiske step.

However, we have to notice that our theoretical analysis is
devoted to \emph{averaged quantity } only. As has been shown in
Refs. \cite{Fist1,JETP93}, in inhomogeneous Josephson junctions
strong mesoscopic fluctuations of various physical quantities
occur. These fluctuations manifest themselves in the form of
random oscillations on the dependence of $I_c(n_A)$ (or
$I_c(B_{\parallel})$). These oscillations can be especially large
for the dependence $I^{(1),max}_F$ on $n_A$ due to a strong
nonlinearity of Eqs. (\ref{nonlinampl}) and
(\ref{Joscurrent-gen}).

The pronounced deviation of the experimental point at $n_A=0.15~
{\mu m}^{-2} (B_{\perp}= 3.2 ~G)$ and $n_A=0.26 ~{\mu m}^{-2}
(B_{\perp}= 5.3 ~G)$ from the theoretical prediction (see Fig.7)
can be  also attributed to the trapping of so called "monopole"
Abrikosov vortex. The "monopole" vortex penetrates only a one
superconducting electrode of the tunnel junction and its magnetic
field spreads over the total
 area of a "small" Josephson junction. Such Abrikosov vortex trapped near
the junction center, alters the $I_c(B_{\parallel})$ dependence
from the Fraunhofer pattern to the one with two primary lobes and
with suppressed Josephson critical current at $B_{\parallel}=0$.
In addition, the ampiltude of the first Fiske step is maximized.
Even if a large number of Abrikosov vortices is trapped, the
effect of the monopole vortex, pinned in the central part of
junction area, dominates \cite{SATT9}. Because of a statistical
character of the trapping process, the cooling in the fields
$B_{\perp}=3.2~ G$ and $B_{\perp}=5.3 ~G$ can provide the trapping
of "monopole" Abrikosov vortex near the junction center, which, in
turn, maximize the first Fiske step, similar to the magnetic flux
trapping effect in case of a Josephson junction with annular
geometry \cite{APL}.  The shapes of the $I_c(B_{\parallel})$
dependencies shown in Fig. 5c and 5e (especially the curve $e$),
permit to sustain our explanation of the deviation of the
experimental data from theoretical prediction presented in Fig.7.

\section{\label{sec:level7}Conclusions}

We have developed a theoretical and experimental study  of the
Fiske step resonances in low-dissipative Josephson junctions with
randomly distributed misaligned Abrikosov vortices.  This analysis
is an extension of the approach elaborated in Ref. \cite{ref5} to
inhomogeneous Josephson junctions. The local inhomogeneous
magnetic field of misaligned Abrikosov vortices leads to an
additional contribution to the Josephson phase difference. The
influence of randomly distributed misaligned Abrikosov vortices
was described in terms of the correlation area $\sigma$ (see, Eq.
(\ref{Corrarea})) which depends both on the Abrikosov vortex
density $n_A$ and on the misalignment length of vortices $\delta$.
We obtained a peculiar regime as $\sigma~\geq~\sigma^*=S (\frac{
\lambda_J^2 \gamma \omega_0 } { c_0^2} )^2$ characterized by a
weak non-monotonic dependence of the maximum Fiske step amplitude
on $n_A$. In this regime the amplitude of excited electromagnetic
waves in the junction $a_n \geq 1$. This behaviour is in a sharp
contrast with a strong decrease of the maximum critical current
$I_c$ with $n_A$. The critical parameter $\sigma^*$ depends on the
dissipation $\gamma$ in the junction. For low-dissipative
junctions the parameter $\sigma^*<<S$.

We carried out the experiments on small low-dissipative Nb based
Josephson tunnel junction with artificially introduced Abrikosov
vortices. We measured both the critical current $I_c$ and the
Fiske resonant steps in the $I-V$ curves. We observed that the
critical current $I_c$ strongly decreases  with the density of
Abrikosov vortices in a complete agreement with a theory, Eq.
(\ref{Criticalcurrent-1}).

In the absence of trapped Abrikosov vortices the experimental
dependence of the amplitude of first Fiske resonance on
$B_{\parallel}$ is in a good agreement with the theoretical
description by Eqs. (\ref{nonlinampl}) and (\ref{Joscurrent-gen}),
where the term $J_1$ is taken into account.
As the misaligned Abrikosov vortices were introduced in the
junction, Fiske resonances were observed \emph{even} in the
absence of $B_{\parallel}$. However, an application of a small
magnetic field $B_{\parallel}$ allowed to increase the amplitude
of Fiske resonances. The voltage position of the first Fiske
resonance determined by a geometry of the junction, did not vary
in the presence of Abrikosov vortices. The measured dependence of
the maximum amplitude of first Fiske step $I^{(1),max}_F$ on $n_A$
showed the qualitative agreement with our theoretical analysis
(Eq. (\ref{IF-large})) i.e. with a weak non-monotonic dependence
(see Fig. 7). Therefore, the regime $\sigma \geq \sigma^*$ for our
low-dissipative junctions was realized.

Finally, we notice that Eq. (\ref{IF-large}) describes the
\emph{averaged} value of the amplitude of Fiske resonances.
However, as  was indicated in Refs. \cite{Fist1,JETP93}, strong
mesoscopic fluctuations of various physical values can be observed
in inhomogeneous Josephson junctions. These mesoscopic
fluctuiations are especially strong for the amplitude of Fiske
resonances due to nonlinear character of Eqs. (\ref{nonlinampl})
and (\ref{Joscurrent-gen}). Therefore, observed deviations of some
experimental point from the theoretical curve described by
Eq.(\ref{IF-large}), can be attributed to such mesoscopic
fluctuations. This discrepancy could be also due to a possible
trapping of a single \emph{monopole} Abrikosov vortex just near
the junction center.

\textbf{Acknowledgments}

The authors thank R. Cristiano, S. Pagano and L. Frunzio for the
substantial contribution to the experiment, G. Giuliani and
C.Nappi for the helpful theoretical discussion. M. V. F.
acknowledges the financial support by SFB 491. The final stage of
the preparation of the paper was inspired by the Experiment «
SUPERGAMMA » supported by the Italian Institute of Nuclear Physics
(INFN).


\end{document}